\documentstyle[12pt]{article}
\def\s#1{   \theta_{ #1}  }

\def\t#1{   \theta_{\tilde #1}  }
\def\st#1#2{   \theta_{{ #1}{\tilde #2}}   }
\def\tt#1#2{   \theta_{{\tilde #1}{\tilde #2}}   }
\def\tt#1#2{   \theta_{{\tilde #1}{\tilde #2}}   }

\def\ttt#1#2#3{   \theta_{{\tilde #1}{\tilde #2}{\tilde #3}}   }
\begin{document}
\begin{titlepage}
\begin{centering}
\vspace*{1cm}
{\large{\bf Radiative Electroweak symmetry breaking
 in the MSSM and Low Energy Thresholds }}   \\
\vspace*{6.0ex}
{\large{\rm\bf{K. Tamvakis\footnote[1]{Talk presentend in
 Susy '95, Ecole Polytechnique, Paris, May 1995}}}}   \\
\vspace*{1.5ex}
{\large\it{Division of Theoretical Physics,
University of Ioannina
\\
Ioannina, GR - 451 10,
GREECE}}\\
\vspace*{4.5cm}
{\bf Abstract}\\
\vspace*{3.0ex}
\end{centering}
{\noindent
We study Radiative Electroweak Symmetry Breaking in the Minimal Supersymmetric
 Standard Model (MSSM).
We employ the 2-loop Renormalization Group equations for running masses
 and couplings taking into
account sparticle threshold effects. The decoupling of each
particle below its threshold is realized by  a step function in all one-loop
Renormalization Group equations (RGE). This program
requires the calculation of all wavefunction, vertex  and mass
renormalizations for all particles involved.
Adapting our numerical routines to take care of the succesive decoupling of
each particle below its threshold, we compute the mass spectrum of sparticles
and Higgses consistent with the existing experimental constraints.
The effect of the threshold corrections is in general of the same order of
magnitude  as the two-loop contributions with the exception of the
heavy Higgses. }
\end{titlepage}

\par

The purpose of the present talk is to report briefly on a treatment
of low energy threshold effects in the Renormalization Group equations
 of the parameters of the MSSM in the framework of Radiative Electroweak
breaking. Since we have employed
the
${\overline{DR}}$ scheme in writting down the one-loop Renormalization
 Group
 equations, which is by definition mass-indepedent, we could ``run"
them from
$M_{X}$ down to $M_{Z}$ without taking notice of the numerous
sparticle
thresholds existing in the neighborhood of the supersymmetry breaking
 scale
near and above $M_{Z}$. This approach of working in the ``full"
theory
consisting of particles with masses varying over 1-2 orders of
 magnitude has
to overcome the technical problems of the determination of the pole
 masses. Our
approach, also shared by other analyses, is to introduce a succession
 of
effective theories defined as the theories resulting after we
functionally
integrate out all heavy degrees of freedom at each particle threshold.
 Above
and below each physical threshold we write down the Renormalization
Group
equations in the ${\overline{DR}}$ scheme only with the degrees of
freedom
that are light in each case. This is realized by the use of a theta
function at
each physical threshold. The integration of the Renormalization Group
equations
in the ``step approximation" keeps the logarithms $\ln(\frac{m}{\mu})$
 and
neglects constant terms. The physical masses are determined by the
condition
$m(m_{phys})=m_{phys}$ which coincides with the pole condition if we
keep
leading logarithms and neglect constant terms. The great advantage of
 this
approach is that the last step of determining the physical mass
 presents no
extra technical problem and it is trivially incorporated in the
integration of
the Renormalization Group equations.

A dramatic simplification of the structure of the supersymmetry
 breaking
interactions is provided either by Grand Unification assumptions or
 by
Superstrings.
The simplest possible choice
at tree level
is to take all sparticle and Higgs masses equal to a common mass
 parameter
$m_{o}$, all gaugino masses equal to some parameter $m_{1/2}$ and
all cubic
couplings flavour blind and equal to $A_{o}$. This situation is
 common in the
effective Supergravity theories resulting from Superstrings but there
 exist
more complicated alternatives. For example Superstrings with massless
 string
modes of different modular weights lead to different sparticle masses
at tree
level$^{\cite{matalliotakis}}$. The equality of gaugino masses can also
 be circumvented in an
effective supergravity theory with a suitable non-minimal gauge kinetic
 term$^{\cite{ellis}}$. Note
however that such non-minimal alternatives like flavour dependent
sparticle
masses are constrained by limits on FCNC processes. In what follows
 we shall
consider this simplest case of four parameters $m_{o}$, $m_{1/2}$,
 $A_{o}$ and
$B_{o}$.
The scalar potential of the model is
\begin{eqnarray}
V &=& m_{1} ^{2} |H_{1}|^{2} +m_{2} ^{2} |H_{2}|^{2} +
\mu B (H_{1} H_{2} +c.c) \nonumber \\
&+& {\frac {1} {8}} g'^{2} (|H_{1}|^{2} -|H_{2}|^{2})^{2} +
{\frac {1} {8}} g^{2} (|H_{1}|^{4} +|H_{2}|^{4} +
4 |H_{1} ^{\dagger} H_{2}|^{2}
-2 |H_{1}|^{2} |H_{2}|^{2}) +.....
\end{eqnarray}
written in terms of
\begin{equation}
   m_{1,2} ^{2} \equiv m_{H_{1,2}} ^{2} + {\mu}^{2}
\end{equation}
Only an unbroken minimum appears for $m_{1} ^{2}=m_{2} ^{2}$.
Replacing the
appearing parameters with their running values
$m_{1} ^{2} (Q)$, $m_{2} ^{2}
(Q)$,... as defined by the Renormalization Group and adding
the one-loop
radiative corrections obtained in the $\overline{DR}$ scheme,
\begin{equation}
\Delta V_{1} = {\frac {1} {64 { \pi}^{2}}} Str\{{\cal M}^{4} (ln(
{\cal M}^{2}/ {Q^{2}})-3/2)\} \end{equation}
we end up with an Effective Potential that upon minimization
supports a vacuum
with spontaneously broken electroweak
symmetry$^{\cite{castano}\cite{zwirner}}$. A reasonable
approximation to (3) would be to allow only for the
dominant top-stop loops. Note that although the
Renormalization Group improved tree level potential
depends on the scale Q
this is not the case for the full 1-loop Effective
Potential which is
Q-independent up to,  irrelevant for minimization, Q-dependent but
field-independent terms.

We shall assume that at a very high energy scale $M_{X}$ the soft
supersymmetry
breaking is represented by four parameters $m_{o}$, $m_{1/2}$,
$A_{o}$ and $B$
of which we shall consider as input parameters only the first
three and treat
$B(M_{Z})$ as determined by minimization conditions of the one loop effective
potential. Actually we can treat $\beta(M_{Z})$ as input parameter and both
$B(M_{Z}), \mu(M_Z)$ are determined by solving the minimization conditions
 with the sign of $\mu$ left undetermined.
The top-quark mass$^{\cite{abe}}$, or
equivalently the top-quark Yukawa coupling, although localized
 in a small range
of values should also be considered as an input parameter since
the sparticle spectrum and the occurrance of symmetry breaking
itself is
sensitive to its value.Thus, the input parameters are $m_{o}$,
$m_{1/2}$, $A_{o}$, $\beta (M_{Z})$ and $m_{t} (M_{Z})$ as well as the sign of
$\mu$.

In our notation, for a physical mass $M$,
\begin{equation} \theta_{M} \equiv \theta (Q^{2} -M^{2})
\end{equation}
Also $t$ stands for $t=ln Q^{2} $ and $\beta_{\lambda} \equiv {\frac
{d{\lambda}} {dt}}$ for each parameter $\lambda$. Note also that we
assume
diagonal couplings in family space.

As an example consider the one loop $\beta$-function of the trilinear
coupling$^{\cite{tamvakis}}$ $A_{\tau}$ ,
\begin{eqnarray}
 {\frac {dA_\tau}{dt} }&=&{\frac {1}{{(4\pi)}^2} }   \{
-3{g_2}^2 {M_2} \tt{W}{H_1} -{3 \over 5}{g_1}^2 {M_1}
{(2+\t{H_1})\t{B}}  \nonumber  \\  \nonumber  \\
&+&3{Y_b}^2 {A_b} \tt{D}{Q}+4{Y_\tau}^2 {A_\tau}+{A_\tau}[ {Z_{\tau 1}}
{g_1}^2
+ {Z_{\tau 2}}{g_2}^2 +{Z_{\tau \tau}}{Y_\tau}^2]    \}   \\  \nonumber
\end{eqnarray}

Where $Z_{\tau 1}$, $Z_{\tau 2}$ and $Z_{\tau \tau}$ are displayed in Table I.

\begin{center}
\begin{tabular}{l}\hline
\multicolumn{1}{c}{\bf TABLE I} \\ \hline \\
${Z_{\tau 1}}={3 \over 40} [ 11 +
10\t{B}-8\t{E}-4\tt{B}{E}+8\ttt{B}{E}{H_1}+ $  \\ \\
\hspace{5cm}  $  2\s{H_1}
 -8\st{H_1}{E}-2\t{L}-\tt{B}{L} -8\tt{E}{L}-4\ttt{B}{H_1}{L}
+4\st{H_1}{L}  ]  $
\\  \\
${Z_{\tau 2}}={1 \over 8} [ -3 +
6\s{H_1}-6\t{L}-12\st{H_1}{L}+6\t{W}-3\tt{L}{W}+12\ttt{W}{H_1}{L}] $
\\ \\
${Z_{\tau \tau}}={1 \over 4} [ -16 +
+6\t{H_1}-\tt{H_1}{E}-3\s{H_1} +4\st{H_1}{E} +4\tt{E}{L}
-2\tt{H_1}{L}+ 8\st{H_1}{L}  ]  $ \\ \\ \hline \\
\end{tabular}
\end{center}
\vspace{0.5cm}
\noindent
{\bf Table I}:\quad Threshold coefficients appearing in the renormalization
group equation \\
of the trilinear scalar coupling $A_\tau$.\\

 Note that
the threshold corrections
introduced in our approximation by the theta-functions at 1-loop are
expected to be comparable to the
standard 2-loop RG corrections. In our numerical analysis that we
follow we shall employ the 2-loop RG
equations which have not been presented here due to their complicated
form but can be found
elsewhere$^{\cite{martin}}$.
The problem at hand consists in finding the physical masses of the
presently
unobserved particles, i.e. squarks, sleptons, Higgses, Higgsinos and
gauginos, as well as their physical couplings to other observed
particles. This
will be achieved by integrating the Renormalization Group equations
from a
superheavy scale $M_X$, taken to be in the neighbourhood of $10^{16}
GeV$, down
to a scale $Q_o$ in the stepwise manner stated. If the equation at
 hand is the
Renormalization Group equation for a particular running mass $m(Q)$,
then $Q_o$
is the corresponding physical mass determined by the condition
 $m(Q_o)=Q_o$. If
the equation at hand is the Renormalization Group equation for a
coupling the
integration will be continued down to $Q_o=M_Z$. Acceptable solutions
 should
satisfy the minimization conditions at $M_Z$, i.e. describe a
low energy
theory with broken electroweak symmetry at the right value of
 $M_Z \simeq
91.187 GeV$.

The boundary condition at high energy will be chosen as simple as
possible, postponing for elsewhere the study of more complicated
alternatives. Thus at the (unification) point $M_X$, taken to be
 $10^{16}$ GeV, we shall take
\begin{eqnarray}
m_{\tilde{Q}} (M_X)& =& m_{\tilde{D^c}}
(M_X) = m_{\tilde{U}^c} (M_X) =
m_{\tilde{L}} (M_X) = m_{\tilde{E}^c} (M_X)\nonumber
\\ & =& m_{H_1} (M_X) =
m_{H_2} (M_X) \equiv m_o
\end{eqnarray}
and
\begin{equation}
M_1 (M_X) = M_2 (M_X) = M_3 (M_X) \equiv m_{1/2}
\end{equation}
In addition we take equal cubic couplings at $M_X$, i.e.
\begin{equation}
A_e (M_X) = A_d (M_X) = A_u (M_X) \equiv A_o
\end{equation}

Our set of constraints includes the low energy experimental gauge
coupling values  which we have taken
to be $M_Z =91.187 Gev$ ,
${\alpha (M_Z)^{-1}}_{\overline
{MS}}=127.9\pm 0.1$ and  $(sin^{2}{\theta_W})_{\overline
 {MS}}=0.2316-.88 10^{-7} ({M_t}^2-160^2) Gev^{-2}$. The knowing,
average experimental value of $\alpha_{3}$ is $0.117 \pm 0.010$.
 These $\overline {MS}$ values for the couplings
are related to the relation,  $\overline {DR}$\footnote[1] {Note that at the
 2-loop order the $\overline {DR}$
scheme needs to be modified so that no contribution to the scalar
 masses due to
the``$\epsilon$-scalars"$^{\cite{jack}}$ shows up.}ones through the
relations
$g_{\overline {MS}}=g_{\overline{DR}}(1- C g^2/{96 \pi^{2}})$,
where $C=0,2,3$ respectively for the three factor gauge groups.
For the b-quark and $\tau$-lepton masses we have taken
$m_b=5.0$ Gev and $m_{\tau}=1.8$ Gev.The recent evidence$^{\cite{abe}}$
for the top quark mass has motivated values in the neighborhood of
$176 \pm 8$ Gev.The physical top quark mass $M_t$ is related to the
running top-quark mass through the approximate relation
\begin{equation}
m_t (M_t)=\frac {M_t} {(1+ \frac {5 \alpha_3} {3 \pi} +....)}
\end{equation}

As stated previously the $B,\mu$ are not inputs in the approach we are
following but are determined through the equations  minimizing the scalar
potential. For their determination at the scale $M_Z$ we take into account
the one loop corrected potential considering the dominant top and stop
contributions. This procedure modifies the tree level values $B(M_Z)$,
$\mu(M_Z)$
. It is well known that the value of $\mu$ affects the predictions for the
physical masses especially those of the neutralinos and charginos. In
approaches in which the effect of the thresholds is ignored in the RGE's the
determination of $B,\mu$ is greatly facilitated by the near decoupling of
these parameters from the rest of the RGE's. However with the effects of the
thresholds taken into account such a decoupling no longer holds since the
thresholds themselves depend on $B,\mu$, or equivalently on $\mu,m_3^2$.
Thus, as initial inputs for $B(M_Z)$ and $\mu(M_Z)$ we take those arising
from the       minimization equations assuming that theshold effects are
absent. At this  stage our analysis is identical to those of other authors.
Subsequently we run our numerical routines switcing on the threshold
contributions to  the RGE's keeping fixed the inputs for $A_o$,$m_o$,
$m_{1/2}$, $\tan \beta$ and all couplings. This procedure corrects the
initial inputs for $B(M_Z)$, $\mu(M_Z)$ in each run until convergence is
reached. This is unecessary of course in cases where the
thresholds are neglected. The next step regarding the mixing parameters
$\mu,m_3^2$ is to correct them taking into account the one loop effective
potential in the way prescribed earlier.

We have displayed some of our results in tables II and III.
We have taken $\mu(M_Z)$    positive
and hence $B(M_Z)$  negative. Their mirror values
$\mu(M_Z)<0$, \, $B(M_Z)>0$ lead to qualitatively similar results.
In the table II, for a characteristic set of values
$A_o=400$ GeV, $m_o=300$
GeV and $m_{1/2}=200$ GeV we have varied $tan\beta$ between $2$ and
$25$.  Note the well
 known$^{\cite{nanopoulos}}$
approximate equality between the masses of one of the neutralinos and
one of the charginos. The lightest Higgs turns out to be heavier than the
$Z$ - boson. Althought not displayed, for negative $\mu$ its mass drops
 below $M_Z$ for small values of the angle $tan\beta \simeq 2$.

Finally, Table III compares for a characterestic choice of parameter values,
three dinstict cases.Case [a] indicates one loop predictions, case [b] two loop
predictions with thresholds only in couplings, and case [c] the complete two
loop with thresholds everywhere.Comparison of the first two cases [a] and [b]
point out the fact that thresholds in couplings affect only by small ammount
$(1\%-2\%)$ the spectra, except from the neutralino and chargino states, where
the differences are of order $10\%$  due to different evolutions of the soft
gaugino masses $M_{1,2}$.

Comparing cases [b] and [c], we observe quite large effects in states labeled
as
${\tilde{\chi}}_{3,4}^{o}$,  ${\tilde{\chi}}_{1}^{c}$ as well as in heavy Higgs
states.This discrepancy is due mainly to the
evolution of $m_{3} ^2$, whose values affect substantially the masses of the
pseudoscalar and charged Higgses,and in particular on it's dependense on the
gaugino masses$^{\cite{dedes}}$.


\newpage
\begin{center}
\begin{tabular}{cccccc}\hline
\multicolumn{6}{c}{\bf TABLE II} \\ \hline
\multicolumn{6}{c}{$m_t=175 \: , \:  A_o=400 \: , \: m_o=300 \: , \:
m_{1/2}=200 \: , \: \mu(M_Z) >0$}
\\
$tan\beta$         &25       &20      &15       &10       &2
\\ \hline \\
$M_{GUT}$          & 2.632   &2.633   &2.633    &2.632    &2.515    \\
$(10^{16} GeV)$    &         &        &         &         &         \\
$\alpha_{GUT} $    &.04161    &.04161   &.04161    &.04161    &.04134   \\
$\alpha_{em}^{-1}$ &127.9    &127.9   &127.9    &127.9    &127.9    \\
$sin^{2}{\theta_W}$&.2311    &.2311   &.2311    &.2311    & .2311   \\
$\alpha_{3}$       &.13128   &.13129  &.13129   &.13126   &.12909   \\ \hline
\\
$M_t$              &177.0   &177.0    &177.0    &177.0    &176.8
\\ \hline   \\
$\tilde{g}$        &495.6   &495.8    &495.9    &495.9   &492.4     \\ \\
${\tilde{\chi}}_{1}^{o}$   &77.1      &77.0    &76.8    &76.4   &74.7    \\
${\tilde{\chi}}_{2}^{o}$   &139.1     &138.9   &138.5   &137.6  &136.0   \\
${\tilde{\chi}}_{3}^{o}$   &350.5     &352.3   &354.7   &359.1  &487.8   \\
${\tilde{\chi}}_{4}^{o}$   &-337.2    &-338.6  &-340.3  &-343.2 &-467.7  \\
\\
${\tilde{\chi}}_{1} ^{c}$  &352.3     &354.0   &356.0   &359.9  &484.5  \\
${\tilde{\chi}}_{2} ^{c}$  &138.8     &138.6   &138.1   &137.0  &134.7
\\  \hline \\
${\tilde{t}}_1$,${\tilde{t}}_2$  &527.3,315.4   &531.9,315.2   &535.9,314.4
				 &539.4,312.3   &543.9,285.3   \\
${\tilde{b}}_1$,${\tilde{b}}_2$  &506.2,441.0   &514.2,451.4   &520.9,459.4
				 &525.8,465.0   &525.9,462.4   \\ \\
${\tilde{\tau}}_1$,${\tilde{\tau}}_2$
				 &328.0,266.5   & 330.4,281.0  &331.7,292.9
				 &331.9,302.1   &329.4,309.2       \\
${\tilde{\nu}}_{\tau}$           &306.4         &311.5         &315.6
				 &318.6         &323.4       \\ \hline \\
${\tilde{u}}_{1,2}$,${\tilde{u}}_{1,2}^{c}$
				 &537.8,528.9  &537.8,528.9    &537.8,528.9
				 &537.7,528.8  &535.4,525.8      \\
${\tilde{d}}_{1,2}$,${\tilde{d}}_{1,2}^{c}$
				 &543.3,529.6  &543.3,529.6    &543.3,529.6
				 &543.2,529.5  &538.6,525.8      \\ \\
${\tilde{e}}_{1,2}$,${\tilde{e}}_{1,2}^{c}$
				 &330.1,311.7  &330.1,311.7   &330.0,311.7
				 &330.0,311.6  &328.8,310.3       \\
${\tilde{\nu}}_{1,2} $
				 &321.0        &321.0         &320.9
				 &321.0        &323.5      \\ \hline \\
$A$                &630.7        &612.0        &588.1   &560.1    &669.0  \\
$h_o$,$H_o$        &114.1,630.6  &114.2,611.9  &114.2,588.0
		   &113.8,560.2  &92.1,672.8                              \\
$H^{\pm}$          &635.4        &616.9        &593.1   &565.4     &673.5
\\ \hline
\end{tabular}
\end{center}

\vspace{1.cm}
\noindent
{\bf Table II}:\quad MSSM predictions for $m_t=175 \,GeV,\,
A_o=400 \, GeV$,$\,m_o=300$,
\,${m_{1/2}}=200 GeV$ and for values of $\tan \beta$ ranging from 2 to 25.
Only the $\mu > 0$ case is displayed.
\newpage
\begin{center}
\begin{tabular}{cccc}\hline
\multicolumn{4}{c}{\bf TABLE III} \\ \hline
\multicolumn{4}{c}{$m_t=175$, $tan\beta=10$, $A_o=250$, $m_o=200$,
$m_{1/2}=150$, $\mu(M_Z) >0$} \\ \hline
	  & Case [a]    &  Case [b]        & Case [c]             \\
	  & 1-loop      &  2-loop          & Complete 2-loop        \\
	  &     &(thresholds in couplings)  & \\
\hline
$M_{GUT}$          & 2.1881    &2.8876        &2.8766           \\
$(10^{16} GeV)$    &                  &             &         \\
$\alpha_{GUT} $    &.04127     &.04201       &.04202          \\
$\alpha_{em}^{-1}$ &127.9             &127.9        &127.9    \\
$sin^{2}{\theta_W}$&.23105            &.23110       &.23110  \\
$\alpha_{3}$       &.11767    &.13284        &.13289       \\ \hline
\\
$M_t$              &181.0             &177.0         &177.0
\\ \hline   \\
$\tilde{g}$        &398.4      &381.4       &382.6            \\ \\
${\tilde{\chi}}_{1}^{o}$   &59.2  &55.0   &54.4           \\
${\tilde{\chi}}_{2}^{o}$   &109.0  &98.3  &96.7           \\
${\tilde{\chi}}_{3}^{o}$   &302.8     &304.1  &279.6       \\
${\tilde{\chi}}_{4}^{o}$   &-284.1    &-287.7   &-260.1       \\
\\
${\tilde{\chi}}_{1} ^{c}$  &304.0        &305.5   &280.9       \\
${\tilde{\chi}}_{2} ^{c}$  &108.0   &97.4       &95.3
\\  \hline \\
${\tilde{t}}_1$,${\tilde{t}}_2$  &443.6,247.0   &442.0,235.2    &440.2,234.7
  \\
${\tilde{b}}_1$,${\tilde{b}}_2$  &401.7,357.2  &395.4,352.8   &394.9,352.6
   \\ \\
${\tilde{\tau}}_1$,${\tilde{\tau}}_2$
				 &235.3,203.6 &232.2,201.6  &231.3,202.7
   \\
${\tilde{\nu}}_{\tau}$           &216.6       &212.6     &212.5
   \\ \hline \\
${\tilde{u}}_{1,2}$,${\tilde{u}}_{1,2}^{c}$
			       &410.5,401.4   &400.1,394.1  &400.1,394.1
  \\
${\tilde{d}}_{1,2}$,${\tilde{d}}_{1,2}^{c}$
			       &417.8,402.4 &407.5,395.7   &407.5,395.7
  \\ \\
${\tilde{e}}_{1,2}$,${\tilde{e}}_{1,2}^{c}$
			       &231.6,212.6  &227.7,211.6  &227.5,211.8
  \\
${\tilde{\nu}}_{1,2} $
				 &218.1      &214.2     &214.1
  \\ \hline \\
$A$                       &412.1       &421.0        &394.5           \\
$h_o$,$H_o$               &113.0,412.2 &110.7,421.1  &110.5,394.7     \\
$H^{\pm}$                 &419.5       &428.1        &402.1
\\ \hline
\end{tabular}
\end{center}
\vspace{1.cm}
\noindent
{\bf Table III}:\quad MSSM mass spectrum for the inputs shown in the first
row $(\mu>0)$. We compare 1 - loop (case [a]), 2 - loop with thresholds in
couplings (case [b]) and complete  2 - loop predictions (case [c]) with
thresholds in both couplings and dimensionful parameters.






\begin{thebibliography}{99}
\bibitem{castano}
G. Gamberini, G. Ridolfi and F. Zwirner, Nucl. Phys. B331(1990)331;

R. Arnowitt and P. Nath, Phys. Rev. D46(1992)3981;

D. J. Casta\~{n}o, E. J. Piard and P. Ramond, Phys. Rev. D49(1994)4882.
\bibitem{matalliotakis}
D. Matalliotakis and H.-P. Nilles, Nucl.Phys.B435(1995)115;

A. Lleyda and C. Munoz, Phys.Lett.B317(1993)82;

N. Polonsky and A. Pomarol, Phys.Rev.D51(1995)6532;

Ph. Brax, U. Elwanger and C. A. Savoy, Phys.Lett.B347(1995)269;

J. Louis and Y. Nir, hep-ph/9411429.
\bibitem{ellis}
J. Ellis, K. Enqvist, D. V. Nanopoulos and K. Tamvakis,
Phys. Lett. B155(1985)381.
\bibitem{zwirner}
J. Ellis, G. Ridolfi and F. Zwirner, Phys. Lett. B257(1991)83 , B262(1991)477;

Y. Okada, M. Yamaguchi and T. Yanagida, Prog. Theor. Phys. 85(1991)1;

H. E. Haber and R. Hempfling, Phys. Rev. Lett. 66(1991)1815;

A. Brignole, Phys. Lett. B277(1992)313, Phys. Lett. B281(1992)284;

M. Drees and M. M. Nojiri, Phys. Rev. D45(1992)2482.
\bibitem{abe}
F. Abe et.al, Phys.Rev.Lett. 74(1995)2676;

S. Abach et.al, Phys.Rev.Lett.74(1995)2632.
\bibitem{tamvakis}
A. B. Lahanas and K. Tamvakis, Phys.Lett.B348(1995)451.

``Radiative Electroweak symmetry breaking in the MSSM and \\ low energy
thresholds" A.Dedes, A.B. Lahanas and K.Tamvakis, hep-ph/9504239.
\bibitem{martin}
S. P. Martin and M. T. Vaughn, Phys.Rev. D50(1994)2282;

Y.Yamada, Phys.Rev D50(1994)3537;

I. Jack and D. R. T. Jones, Phys.Lett.B333(1994)372.
\bibitem{jack}
I. Jack, D. R. T  Jones, S. P. Martin, M. T. Vaughn and Y. Yamada,\\
Phys.Rev D50(1994)5481.
\bibitem{nanopoulos}
R. Arnowitt and P. Nath, Phys. Rev. Lett. 69(1992)725,\\
Phys. Lett. B 289(1992)368;

J. L. Lopez, D. V. Nanopoulos and H. Pois, Phys. Rev.D47(1993)2468.
\bibitem{dedes}
A. Dedes, A. B. Lahanas and K. Tamvakis, in preparation.
\end{thebibliography}
\end{document}